\documentclass[prd,aps,a4paper,eqsecnum,nofootinbib,floatfix,twocolumn]{revtex4}  

\newif\ifusesec
\usesectrue  
   
\usepackage{graphicx} 
\usepackage{amsmath,amsfonts,amssymb}

\newcommand{\beq}{\begin{equation}}
\newcommand{\eeq}{\end{equation}}
\newcommand{\bea}{\begin{eqnarray}}
\newcommand{\eea}{\end{eqnarray}}

\begin{document}

\title{First Post-Minkowskian approach to turbulent gravity}

\author{Donato Bini$^{1,2}$, Stuart Kauffman$^3$, Sauro Succi$^{1,4,5}$,  Pablo G. Tello$^6$}
  \affiliation{
$^1$Istituto per le Applicazioni del Calcolo  \lq\lq 
M.~Picone,  \rq\rq 
CNR, I-00185 Rome, Italy\\
$^2$INFN, Sezione di Roma Tre, I-00146 Rome, Italy\\
$^3$Institute for Systems Biology, Seattle, WA 98109, USA\\
$^4$
Istituto Italiano di Tecnologia, 00161 Rome, Italy\\
$^5$Physics Department,
Harvard University, Cambridge,USA\\
$^6$CERN, Geneva, Switzerland\\
}

\date{\today}

\begin{abstract}

We compute the metric fluctuations induced by a turbulent 
energy-matter tensor within the first order Post-Minkowskian approximation.
It is found that the turbulent energy cascade can in principle interfere
with the process of black hole formation, leading to a potentially strong
coupling between these two highly nonlinear phenomena.
It is further found that a power-law turbulent energy spectrum $E(k) \sim k^{-n}$
generates metric fluctuations scaling like $x^{n-2}$,  where $x$ is the   four-dimensional spacelike
distance from an arbitrary origin in  Minkowski  spacetime,
highlighting the onset of metric singularities whenever $n <2$.
Finally,  the effect of metric fluctuations on the geodesic motion of test particles
is also discussed as a potential technique to extract information on the spectral
characteristics of fluctuating spacetime.
\end{abstract}

\maketitle

\section{Introduction}

The key informing principle of general relativity stipulates that matter/energy 
and spacetime {\it co-evolve} through a self-consistent loop, whereby, to
say it with Wheeler,  \lq\lq spacetime tells matter how to move; matter 
tells spacetime how to curve"~\cite{WHEELER}.
 
Despite its logical simplicity,  the mathematical formulation
of the above statement (Einstein equations, EEs for short) faces 
with a daunting complexity barrier,  mostly on account
of the strong non-linearity of the matter-spacetime interactions.  
Even for the \lq\lq simple" case of matter at rest, the exact solutions of the
EEs are restricted to very few precious instances, usually characterized 
by highly idealized geometries with very special symmetry 
properties (often too special), which impair a general 
understanding of the problem \cite{REZZOLLA}.

Evaluating the gravitational field generated by matter in motion
clearly adds another layer of mathematical complexity,  particularly
in the case where such motion is  not regular but {\it turbulent} instead.
For instance,  it is not known what kind of spacetime metric results 
from a given fluctuating energy-momentum tensor 
(energy density, pressure, velocity field).
Likewise,  we do not know the  fate of turbulent flows in the
presence of gravity: do the associated scales (gravitational and turbulent) 
compete or cooperate among them? 
Does gravity always dominate in the end,  erasing,  perhaps beyond some threshold,  
all fluid scales,  or do the latter leave an appreciable long-standing signature on the 
gravitational field despite its dominance? 
In other words: can turbulence play an appreciable role on the gravitational collapse 
process (or in a cosmological context) of a turbulent fluid?

The relevance of these questions for modern astrophysics and cosmology 
cannot be overstated~\cite{Yang:2014tla,Marochnik_I,Marochnik_II,Barreto:2022len,Dahl:2021wyk,Galtier:2021ovg,RoperPol:2021gjc,Waeber:2021xba,Adams:2013vsa}, and this work 
represents a preliminary attempt to gain semi-quantitative insights into the above matters.

More specifically,  we proceed  within a Post-Minkowskian (PM) framework, i.e. starting from a flat space 
situation (zero gravity, and Minkowskian fluid dynamics) and adding corrections 
to the first order in the gravitational constant $G$, eventually to be 
continued with high-order iterative corrections. 
This is a standard approach in the study of the 
two-body problem in general relativity and seems to offer a 
promising avenue also for the case of fluid-driven gravitational field. 
In the following, we shall present a \lq\lq warm-up"
investigation along these lines,  highlighting on the various difficulties which stand on the
way of a quantitative understanding of the turbulence-gravity coupling.  

\section{The turbulent energy cascade and its interference with metric length scales: dimensional estimates}

We begin by considering a gravity-free (flat space) turbulent fluid, whose velocity fluctuations
at statistical steady-state,  obey the following generic power-law statistics:
\begin{equation}
u(L)  = u(L_0) (L/L_0)^{\alpha}\,,
\end{equation}
where $L$ is a generic, running, length scale and $L_0$ is the typical size of the fluid, related to the typical 
velocity size $u(L_0)$, $\alpha$ is a scaling exponent in the range $0 \le \alpha \le 1$, with
$\alpha=0$ corresponding to white uncorrelated noise (total randomness), while 
$\alpha=1$ denotes a smooth,  differentiable field.
In the following we shall refer to $\alpha$ as to the velocity roughness exponent.

\begin{figure}
\centering
\includegraphics[scale=0.45]{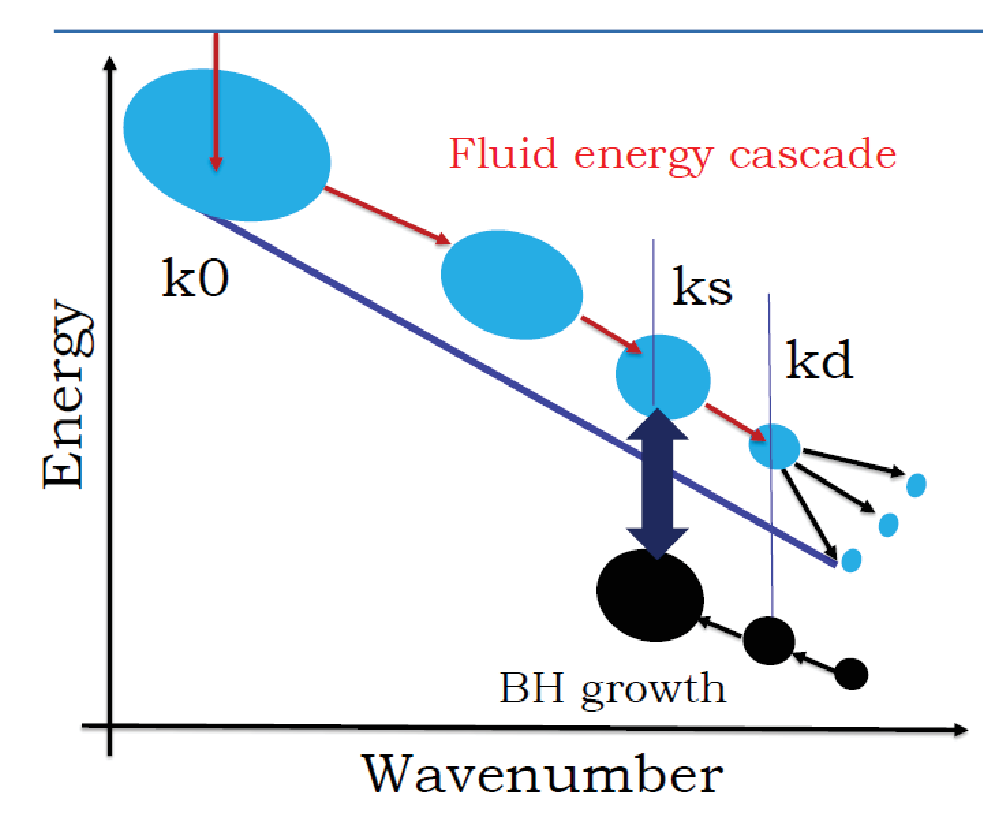} 
\caption{\label{fig:1}At sufficiently high Reynolds numbers,  the energy cascade 
reaches a critical point where dissipative Kolmogorov eddies overlap 
with the length scale associated with the metric background,  in the case 
of the figure, the Schwarzschild  radius of a growing black-hole.
}
\end{figure}

Starting from a mother eddy of size $L_0$,  the
nonlinear cascade generates eddies of progressively smaller size,  till the smallest
active length is reached, below which nonlinearity is no longer capable of
sustaining coherent motion against dissipation.
This happens at the Kolmogorov (or dissipative) length, which
is given by the following expression \cite{K41,ESS,SREENI}
\beq
L_d = \frac{L_0}{{\it Rey}^{1/(1+\alpha)}}\,, 
\eeq
where ${\it Rey}=U_0L_0/\nu$ denotes the Reynolds number
of a turbulent fluid with kinematical viscosity $\nu$ \cite{FRISCH}. 
The situation is schematically represented in Fig. \ref{fig:1}.
Let $L_m=\nu/U_0$ (such that ${\it Rey}=L_0/L_m$) be a microscale length fixed 
by the ratio kinematic of the kinematic viscosity $\nu$ and 
the macroscopic velocity $U_0$ of a  
fluid  of macroscale $L_0$.  
A simple rearrangement leads to the following compact expression:
\begin{equation}
\label{LD}
L_d = L_0^{p} L_m^{1-p} \,,
\end{equation}
where the scaling exponent $p$ relates
to the roughness via  $p=\alpha/(1+\alpha)$;  for example, $0 \le p \le 1/2$, assuming $\alpha\in [0,1]$.

Since  the kinematic viscosity shows surprisingly
small variations across disparate states of matter \cite{Trachenko:2019ghg},  
we  keep it within the range $10^{-4}$
(International System units, ten times air in standard conditions) to $10^{-7}$ (quark-gluon plasma, QGP) through 
the empirical law $L_m \to L_q (\rho_q/\rho)^{1/6}$. Here $L_q \sim 10^{-15}\, m$
and $\rho_q \sim 10^{18}\, Kg/m^3$ are the QGP mean free path 
and density, respectively, while $\rho$ denotes the density of the fluid 
introducing another scale, say the density-related gravitational length
\begin{equation}
L_g = \frac{c}{\sqrt{G \rho}}\,.
\end{equation}

This accounts for three  orders in magnitude change in viscosity 
over eighteen orders of magnitude change in density.

The expression (\ref{LD}) shows that  $L_d$ is an intermediate 
mesoscale ranging from $L_m$ at $\alpha \to 0$ to $L_0$ as 
$\alpha \to \infty$, with $L_d=\sqrt{L_0L_m}$ in the case $\alpha=1$ ($p=\frac12$).
Hence the question is whether, depending on the values of the physical-geometrical
parameters at hand,  there exist reasonable values of the roughness exponent such
that the dissipative length can be made comparable or smaller 
than the typical curvature length of the spacetime, say the 
Schwarzschild scale $L_s = G M_0/c^2$ where  $M_0$ is the mass equivalent of $L_s$.
This is a condition for {\it strong-coupling} between the geometrical background curvature scale  
and hydrodynamic turbulence of the surrounding matter field \cite{K41}.
By demanding $L_d < L_s$, we obtain:

\begin{equation}
\label{STRONG}
L_s  > {L_0}^p L_m^{1-p}\,.
\end{equation}
Replacing $L_0$ in terms of $M_0$, i.e., $L_0=\left(\frac{M_0}{\rho}\right)^{1/3}$,  Eq. \eqref{STRONG} becomes
\beq
M_0>\left( \frac{c^2}{G\rho^{p/3}} \right)^{\frac{1}{1-\frac{p}{3}}}L_m^{\frac{1-p}{1-\frac{p}{3}}}\,.
\eeq
To express this condition in a dimensional form let us divide both sides, for example, by the mass of the Sun, $M_{\rm sun}=2\cdot 10^{30}\, kg$,  $G M_{\rm sun}/c^2=L_{\rm sun}\sim 10^3\, m$ and introduce the gravitational length to replace $c^2/G=L_g^2\rho$. We find the relation
\beq
\label{MSTAR}
m \equiv \frac{M_0}{M_{\rm sun}}>L_g^{a_p} L_m^{1-a_p}L_{\rm sun}^{-1}\equiv m^*\,,
\eeq
where
\begin{equation}
a_p=\frac23 \frac{ p}{1-\frac{p}{3}}\,.
\end{equation}
Here   $m^*$  represents the critical mass above which the dissipative length
falls below the Schwarzschild scale.

Eq. \eqref{MSTAR} can also we written as
\begin{equation}
\label{MSTAR2}
m^* =  \frac{L^*}{L_{\rm sun}}\,,\qquad L^*=L_g^{a_p} L_m^{1-a_p}\,,
\end{equation}
where the numerator defines an effective length, $L^*$, interpolating between
the microscale $L_m$ and the gravitational macroscale $L_g$.
Clearly, $L^*$ is an increasing function of $p$, going from $L_m$ 
at $p=0$ (random fluid) to $L_g^{2/5} L_m^{3/5}$ for $p=1/2$ (smooth fluid).

The expression \eqref{MSTAR} is the main result of this section.
Summarizing, for any value of $p$, we have the following 
length scales related to the density of the fluid
\bea
&& m_*=10^{-3}L_g^{a_p}L_m^{1-a_p}\,,\qquad L_d=L_0^pL_m^{1-p}\,,\nonumber\\ 
&& L_g=\frac{c}{\sqrt{G}}\frac{1}{\rho^{1/2}}=\frac{3.7\cdot 10^{13}}{\rho^{1/2}}\,,\nonumber\\
&& L_m=\frac{10^{-12}}{\rho^{1/6}}\,,\nonumber\\
&& L_0=\left(\frac{M_0}{\rho}\right)^{1/3}=\left(\frac{m_*M_{\rm sun}}{\rho}\right)^{1/3}=10\, m_*^{1/3}L_g^{2/3}\,,\nonumber\\
&& L_s=\frac{GM_0}{c^2}=L_{\rm sun}m_*=10^3 m_* \,.
\eea

Despite their simplicity,  the above expressions invite a number of 
informative remarks.
In the following we analyze three distinguished scenarios of
decreasing roughness, namely:
\begin{enumerate}
\item   Fully random fluid ($\alpha=0$, $p=0$, $a_p=0$,);
\item  Three-dimensional incompressible  fluid  ($\alpha=1/3$, $p=1/4$, $a_p=2/11$);
\item  Two-dimensional incompressible  fluid  ($\alpha=1$, $p=1/2$, $a_p=2/5$).
\end{enumerate}
 
\subsection{Fully random fluid}

In this case we have $\alpha=0$, hence $p=0$, $a_p=0$, and a $|{\mathbf k}|^{-1}$ spectrum.
The expression \eqref{MSTAR} reduces to
\begin{equation}
m^* =   L_m  L_{\rm sun}^{-1} \sim 10^{-3} L_m \,,
\end{equation}
and $L_d=L_m$.
The above relation shows $L_m=L_d=L_s$ varies from $10^{-12}m$ to $10^{-14}m$, while $L_0$ from $10^{5}m$ to $10^{-2}m$.
Moreover the critical mass $m^*$ ranges from 
$10^{-16} \div 10^{-18}$ solar masses, indicating that 
pretty small black-holes can potentially  interfere with a 
turbulent cascade at Reynolds in the order of $10^{16}$. 
Since full randomness is less realistic than correlated
turbulence,  it is of interest to directly inspect the turbulent cases.

\subsection{Three-dimensional turbulence}

As mentioned above, turbulence is a subtly correlated form of chaos
far from pure randomness. As a result, it shows high
sensitivity to spatial dimensionality.  
In $3d$, energy is dissipated even in the (singular) limit of zero viscosity, through the nonlinear
energy cascade from $L_0$ down to $L_d=L_0/{\it Rey}^{3/4}$,  ${\it Rey}$  
being
the Reynolds number. This leads to a roughness exponent 
$\alpha=1/3$ and a $|{\mathbf k}|^{-5/3}$ power spectrum.
Hence, we have $\alpha=1/3$, $p=1/4$, $a_p=2/11$.
The expression  \eqref{MSTAR} now gives
\begin{equation}
m^* = L_g ^{2/11}  L_m^{9/11} L_{\rm sun}^{-1}\,. 
\end{equation}

\begin{figure}
\centering
\includegraphics[scale=0.35]{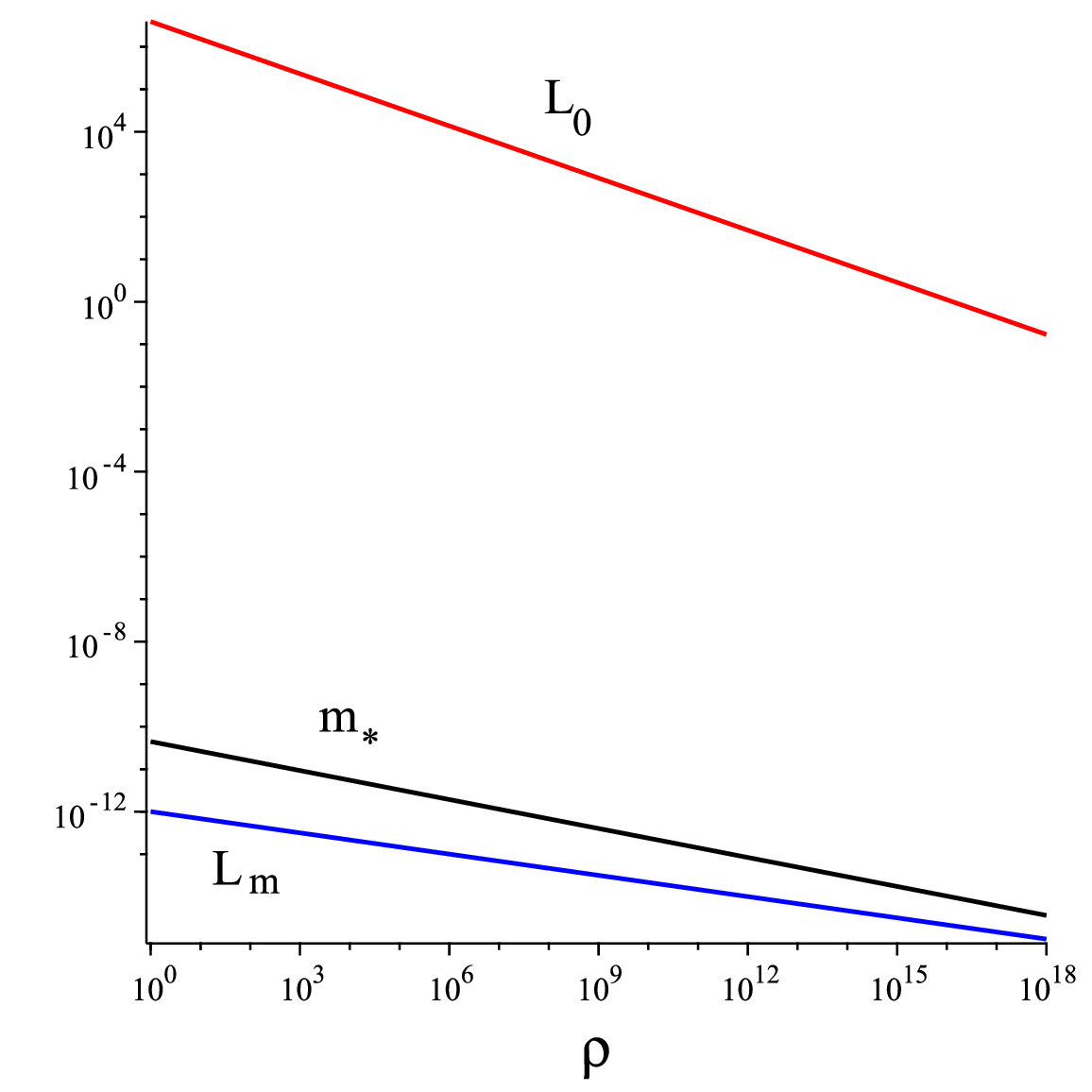}
\caption{\label{fig3} $3d$ Turbulence: $m^*$ 
(black online), $L_0$ (red online),  
and $L_m$ (blue online) are plotted as a function of $\rho$. 
Overlap between gravity and turbulence occurs 
above $10^{-10} \div 10^{-12}$, solar masses, namely
$10^{20} \div 10^{18}$ kg, way more massive than 
the full random case.  The range of Reynolds number is
comparable to the case of full randomness. 
The quantity $m_*$ is dimensionless while $L_0$ and $L_m$ are given in meters.
}
\end{figure}

As an explicit example in Fig. \eqref{fig3}  we show $m^*$, $L_0$  
and $L_m$ as a function of $\rho$. 

The leading factor $L_g$ is now active, but still largely suppressed 
by the small $2/11$ exponent, yet providing a boost of about five orders 
of magnitude with respect to the case of a fully random fluid.
The range of Reynolds numbers (${\rm Rey}=L_0/L_m$) is more less the same 
as for the fully random case.

\subsection{Two-dimensional  turbulence}

In two spatial dimensions,  energy is conserved while enstrophy (vorticity squared)
is dissipated,  which implies a direct (large to small) enstrophy cascade and an
inverse energy cascade (from small to large) \cite{FRISCH}.

The result is a smooth flow field with $\alpha=1$, corresponding to 
$p=1/2$, $a_p=2/5$ and a much steeper energy spectrum $|{\mathbf k}|^{-3}$.
The expression (\ref{MSTAR}) now gives
\begin{equation}
m^* = L_g^{2/5}  L_m^{3/5} L_{\rm sun}^{-1} \,.
\end{equation}
From the above relations   it is apparent that
putative black holes (BHs) are another five orders of magnitude more 
massive than in the $3d$ case. In fact, they 
range from $10^{-5} \div 10^{-9}$
solar masses,  namely $10^{25} \div 10^{21}$ kg, significantly more massive than in the previous cases.

It is worth noting that all turbulent cascades above 
involve pretty large values of the Reynolds number 
${\it Rey}=L_0/L_m$ around $10^{20}$ (as a matter of reference,
the Reynolds number for a standard airline is about $10^8$).

So far we have established that turbulent flows are capable of potentially
strong interactions with growing BHs, since they can reach down 
to the background curvature scale (i.e., the mass in the case of a Schwarzschild black hole). 
The key question, however, is how to describe such strong coupling
in quantitative terms.  As discussed in the Introduction, a fully-fledged 
answer to  this question must necessarily rely upon the non-perturbative
solution of Einstein's equations, driven by a turbulent matter tensor. 

A few heuristic arguments can be brought up, without 
undertaking such a demanding task head-on.
To this regard, let us recall that the standard fate of a dissipative eddy 
of size $L_d$ is to de-cohere into a \lq\lq spray" of droplets too small to 
sustain collective motion: that's where hydrodynamic
bows away to a microscopic description.
Under strong coupling conditions it is plausible to assume that instead 
of being turned into heat, dissipative eddies would rather feed the 
black hole (Primordial Black hole, PBH, could be more appropriate in a cosmological context) growth 
in a sort of preferential way as compared to eddies of larger size.
This speculation is grounded into the principle of locality of turbulence in reciprocal
(Fourier) space, according to which eddies interact strongly only with eddies of comparable 
size.  This is known to hold for fluid turbulence, but does by no means imply that  the
same principle applies to turbulence-gravity interactions as well.

Indeed, recent results, based on the numerical and analytic
solution of the EEs, show that black hole (BH) horizons can themselves turn 
turbulent and in a way which is highly reminiscent of Kolmogorov $3d$ turbulence \cite{K41}.
Such results suggest that the aforementioned principle of locality may indeed
hold true for BH-turbulence interactions as well.
It appears therefore reasonable to speculate that dissipative eddies might
function as \lq\lq catalyzers" of the gravitational collapse, by supplying their
kinetic energy or enstrophy (hence mass) into the newborn BH, long before the entire
mass  
is collapsed. The energy contained in the dissipative eddies is only a
tiny fraction of the total fluid energy, but in the presence of near-singular metrics
even small amounts of extra energy supplies may unleash substantial effects on the
incipient primordial BH dynamics. 
This putative \lq\lq turbulence-assisted" hierarchical gravitational
collapse should be liable to numerical verification.

\section{Turbulence-driven gravity: first order Post-Minkowskian approach}

So far, we have presented statistical steady-state considerations based on dimensional
analysis, an approach which proves exceedingly insightful in the  theory of (gravity-free) turbulence. 
In this section, we endeavor to sketch a quantitative analysis of the EEs
under the stochastic drive of a turbulence matter-energy tensor. 
To this purpose,  we work in a Post-Minkowskian (PM) context which implies 
a weak gravitational field (first order in the gravitational constant $G$, treated 
as a place-holder in the perturbative expansion) but is not restricted 
to small velocities, limiting our considerations to the first-order  
(1PM, or $O(G^1)$) approximation level.

Before plunging into the 1PM formalism, we wish to mention that the effects of stochastic
fluctuations of the energy-matter tensors on the gravitational metric have been considered
before in the context of stochastic gravity \cite{Moffat:1996fu}.  In the approach of Ref. \cite{Moffat:1996fu}  metric fluctuations
are treated by means of a generalized Langevin  formalism \cite{Morozov}, whereas in the present work we 
adopt a strategy inspired by a merger between 1PM and turbulence modeling techniques \cite{TM}.

Let $g_{\mu\nu}=\eta_{\mu\nu}+h_{\mu\nu}$ (with $\eta_{\mu\nu}={\rm diag}[-1,1,1,1]$) denote a 1PM perturbation of the flat space, with inverse $g^{\mu\nu}=\eta^{\mu\nu}-h^{\mu\nu}$ and such that\footnote{At the 1PM order, 
retarded and time-symmetric propagators are equivalent \cite{Damour:2016gwp}.}
\beq
\label{dalemb}
\Box_x h^{\mu\nu}=-16\pi G S^{\mu\nu}+O(\partial \partial hh +h S)\,,
\eeq
where $\Box_x=\eta^{\mu\nu}\partial_\mu\partial_\nu$ and
\bea
S^{\mu\nu}(x)&=&T^{\mu\nu}(x)-\frac12 T(x) g^{\mu\nu}(x)\,,\nonumber\\
T(x)&=&g_{\alpha\beta}(x)T^{\alpha\beta}(x)\,.
\eea
In this case all indices are raised/lowered by using the flat metric $\eta^{\alpha\beta}$, as standard.
Following Ref. \cite{Damour:2016gwp} let us introduce the Green function ${\mathcal G}(x-y)$ of the flat-space D'Alembert operator, such that
\beq
\Box_x {\mathcal G}(x-y)=-4\pi \delta^{(4)}(x-y) \,.
\eeq
This implies
\beq
\label{h_munu_eq}
h^{\mu\nu}(x)=4G \int d^4y {\mathcal G}(x-y)S^{\mu\nu}(y)+O(G^2)\,,
\eeq
and it is fully determined as soon as the source $S^{\mu\nu}(y)$ is specified.
Let us assume $T^{\mu\nu}$ to represent a perfect fluid 
\beq
\label{en_mom}
T^{\mu\nu}=(\rho+p)u^\mu u^\nu+pg^{\mu\nu}\,,\qquad T=-\rho+3p\,,
\eeq
so that
\bea
\label{Smunu}
S^{\mu\nu}&=&(\rho+p)u^\mu u^\nu-\frac12 (p-\rho)g^{\mu\nu}\nonumber\\
&=&\rho \left(u^\mu u^\nu+\frac12 g^{\mu\nu}\right)+p\left(u^\mu u^\nu-\frac12 g^{\mu\nu}\right)\,.
\eea
Furthermore, when studying turbulent motions, characterized by the coexistence of  slowly-varying and rapidly-varying excitations, it is convenient to split the fluid components into an
averaged  and a fluctuating component $X(x)=X_0(x)+\delta X(x)$, e.g.,
\bea
\rho(x)&=& \rho_0(x)  +\delta \rho (x)\,, \nonumber\\
p(x)     &=& p_0(x)  +\delta p (x)\,,\nonumber\\
u^\mu(x)&=&u^\mu_0(x) +\delta u^\mu(x)\,,
\eea
where $X_0(x)$),  is a slowly varying quantity 
whereas $\delta X(x)$ is a rapidly varying component.
In the above "slow (rapid)" implies scales longer (shorter) than the 
typical averaging length, namely the heterogeneity scale of the fluid
(infinity in the case of  homogeneous turbulence).

Within the 1PM  approximation, the source $S^{\mu\nu}$,  being prefactored by $G$, can be treated as a 
zeroth-order quantity i.e., with $g^{\mu\nu}=\eta^{\mu\nu}$ in Eqs. \eqref{Smunu}. 

As a result,  we are left with the following source terms in the equation \eqref{h_munu_eq}:
\beq
S^{\mu\nu}=S_0^{\mu\nu}+S_1^{\mu\nu}+S_2^{\mu\nu}+S_3^{\mu\nu}\,,
\eeq
where
\bea
S_0^{\mu\nu}&=& (\rho_0+p_0)u^\mu_0 u^\nu_0-\frac12 (p_0-\rho_0)\eta^{\mu\nu}\,,\nonumber\\
S_1^{\mu\nu}&=&  (\rho_0+p_0)(u_0^\mu \delta u^\nu+u_0^\nu \delta u^\mu)\nonumber\\
&+& (\delta\rho+\delta p) u_0^\mu  u_0^\nu  -\frac12 (\delta p-\delta \rho) \eta^{\mu\nu}\,,\nonumber\\
S_2^{\mu\nu}&=&  (\rho_0+p_0)R^{\mu\nu}_{u,u}+2u_0^{(\mu}R^{\nu)}_{\rho,u}+2u_0^{(\mu}R^{\nu)}_{p,u}\nonumber\\
S_3^{\mu\nu}&=& R^{\mu\nu}_{\rho, u,u}+R^{\mu\nu}_{p, u,u} \,,
\eea
In the above,  we have  introduced the   \lq\lq correlators"
\bea
R^{\mu\nu}_{u,u} &=&   \delta u^\mu \delta u^\nu \,,\quad
R^{\mu}_{\rho,u} =  \delta \rho \delta u^\mu \,,\quad
R^{\mu}_{p,u} =   \delta p \delta u^\mu \,,\nonumber\\
R^{\mu\nu}_{\rho, u,u}&=& \delta \rho \delta u^\mu \delta u^\nu\,,\qquad
R^{\mu\nu}_{p, u,u}= \delta p \delta u^\mu \delta u^\nu\,,
\eea
where $X_{(ab)}=\frac12(X_{ab}+X_{ba})$ denotes symmetrization.

Notice that $R^{\mu\nu}_{u,u}$ (or more precisely its averaged version) is a 
direct analogue of the Reynolds stress tensor in Kolmogorov 
turbulence,  while $R^{\mu}_{\rho,u}$ and $R^{\mu}_{p,u}$ 
reflect  compressibility effects (in case the pressure is a linear function
of energy  they are basically the same). 

Next, we move to Fourier space,  where the inverse box operator becomes:
\beq
\hat {\mathcal G}(k)=-\frac{1}{k^2 }\,,\qquad k^2=k\cdot k=\eta_{\alpha\beta}k^\alpha k^\beta\,,
\eeq
with $\frac{1}{k^2 }$ denoting a Principal Value (PV) kernel.
Consequently,  
\bea
h_{\mu\nu}(x)&=&-16 \pi G \int \frac{d^4k}{(2\pi)^4}\frac{\hat S_{\mu\nu}(k)}{k^2}e^{ik\cdot x}\nonumber\\
&=& -16 \pi G \sum_{r=0}^3 \int \frac{d^4k}{(2\pi)^4}\frac{\hat S_r{}_{\mu\nu}(k)}{k^2}e^{ik\cdot x}\,,\qquad
\eea
where 
\beq
\hat S_{\mu\nu}(k)=\int d^4x e^{-ik\cdot x}S_{\mu\nu}(x)\,.
\eeq
We can then consider the parts of $h^{\mu\nu}$ sourced by the various components of $S_{\mu\nu}$,
\beq
h_r{}_{\mu\nu}(x)= -16 \pi G  \int \frac{d^4k}{(2\pi)^4}\frac{\hat S_r{}_{\mu\nu}(k)}{k^2}e^{ik\cdot x}\,,\quad r=0,\ldots 3\,.
\eeq
To make further analytical progress,  we need to make physically reasonable 
assumptions on the source terms $\hat S_r{}_{\mu\nu}(k)$, $r=0,1,2,3$. 
For the sake of concreteness, let us start by discussing  the 
case $\hat S_0{}_{\mu\nu}(k)$ (other source terms can be added later), 
under the simplifying hypothesis that $u^\mu_0$ is a constant field, 
\beq
\hat S_0^{\mu\nu}(k)=  \hat \rho_0(k) H_+^{\mu\nu}+ \hat p_0 (k) H_-^{\mu\nu}\,,
\eeq
where 
\beq
H_{\pm}^{\mu\nu}=u^\mu_0 u^\nu_0\pm \frac12 \eta^{\mu\nu}\,,
\eeq
and 
\beq
H_{\pm}^{\mu\nu} u_{0\,\nu}=\lambda_\pm u_0^\mu\,, 
\eeq
with $\lambda_+=-\frac12$ and $\lambda_-=-\frac32$.
Furthermore, for both energy density and pressure we assume a power-law 
scaling,  characteristic of turbulent flows
\beq
\label{hat_rho_p}
\hat \rho_0(k)={\mathcal E}k^n \,,\qquad \hat p_0(k)={\mathcal P}k^m\,,
\eeq
with ${\mathcal E}$ and ${\mathcal P}$ constants, and scaling exponents
$n,m$ both negative.

Finally, let us introduce the notation
\bea
\label{J_q_int}
J_q(x)&=&\int \frac{d^4k}{(2\pi)^4}k^{q}e^{ik\cdot x}\,,
\nonumber\\
&=&\int \frac{d^3k}{(2\pi)^3}e^{ik_ax^a}\int \frac{d\omega}{2\pi}e^{-i\omega t} (-\omega^2+{\mathbf k}^2)^{q/2}\,.\nonumber\\
\eea
For negative values of $q$ this integral is divergent along the lightcones 
$\omega =\pm |{\mathbf k}|$ and should be then regularized (see Appendix A).

We find
\beq
\label{hat_S0_munu}
\hat S_0^{\mu\nu}(k)=   H_+^{\mu\nu} {\mathcal E}k^n +H_-^{\mu\nu} {\mathcal P}k^m \,,
\eeq
and
\bea
h_0^{\mu\nu}(x)&=& -16 \pi G  H_+^{\mu\nu} {\mathcal E}\int \frac{d^4k}{(2\pi)^4} k^{n-2}e^{ik\cdot x}\nonumber\\
&& -16 \pi G H_-^{\mu\nu} {\mathcal P}\int \frac{d^4k}{(2\pi)^4}  k^{m-2}e^{ik\cdot x}\nonumber\\
&=& -16 \pi G  [H_+^{\mu\nu} {\mathcal E}J_{n-2}(x) + H_-^{\mu\nu} {\mathcal P}J_{m-2}(x)]\,.\nonumber\\
\eea
A direct computation shows that, after a Wick rotation and suitable 
regularization techniques (see Appendix A), we obtain:
\beq
\label{J_q_and_C_q}
J_q(x)\to J_q^{\rm Wick}(x)=\frac{C_q}{x^{4+q}}\,,\qquad C_q= \frac{2^q}{\pi^2} \frac{\Gamma(1+\frac{q}{2})}{\Gamma(-(1+\frac{q}{2}))}\,.
\eeq
We note that $C_q$ is well defined In the range $q\in (-2,0)$, $C_q$, with 
values $C_q\in [-\frac1{4\pi^2}, 0]$ ($\lim_{q\to -2}C_q=-\frac1{4\pi^2}$ and $\lim_{q\to 0}C_q=0$).
This implies
\bea
\label{h0_fin_sing}
h_0^{\mu\nu}(x)
&=& -16 \pi G  \left(H_+^{\mu\nu} {\mathcal E}\frac{C_{n-2}}{x^{n+2}} + H_-^{\mu\nu} {\mathcal P}\frac{C_{m-2}}{x^{m+2}}\right)\,,\qquad\nonumber\\
\eea
where $x^2=-t^2+{\mathbf x}^2$ is assumed to be positive, i.e. spacelike (details in Appendix \ref{app:Jq}).
Note that the result \eqref{h0_fin_sing} carries a coordinate-dependent information 
(however, it depends on the choice of the origin of the coordinate system, which we 
implicitly placed at $x=0$). 

This simple example shows the onset of a 
metric singularity (along the lightcone) for $n>-2$ (and same for $m$).
In particular, $2d$ turbulence, $n=-3$,  yields a smooth metric,  while
$3d$ turbulence, $n=-5/3$ leads to a mildly singular one, with the same 
exponent as turbulent velocity fluctuations, but opposite sign, i.e. $-1/3$.
Next, let us examine the $S_1^{\mu\nu}$ term,
\bea
\hat S_1^{\mu\nu}(k)&=& u_0^\mu \int d^4x e^{-ik\cdot x} (\rho_0+p_0)\delta u^\nu\nonumber\\ 
&+&
u_0^\nu \int d^4x e^{-ik\cdot x} (\rho_0+p_0)\delta u^\mu \nonumber\\
&+&H_+^{\mu\nu}\int d^4x e^{-ik\cdot x} \delta \rho(x)\nonumber\\
&+&
H_-^{\mu\nu}\int d^4x e^{-ik\cdot x} \delta p(x)\,.
\eea
The Fourier transform of $\rho_0$, $p_0$, $\delta \rho$ and $\delta p$ leads  to the following expression
\bea
\hat S_1^{\mu\nu}(k)&=& u_0^\mu \int \frac{dk'}{(2\pi)^4}[\hat \rho(k')+\hat p(k')]\hat \delta u^\nu (k-k')\nonumber\\
&+&
u_0^\nu \int \frac{dk'}{(2\pi)^4}[\hat \rho(k')+\hat p(k')]\hat \delta u^\mu (k-k') \nonumber\\
&+&H_+^{\mu\nu}\hat \delta \rho(k)+
H_-^{\mu\nu}\hat\delta p(k)\,.
\eea
Using  Eq. \eqref{hat_rho_p}, we find
\bea
\hat S_1^{\mu\nu}(k)&=& u_0^\mu \int \frac{dk'}{(2\pi)^4}[{\mathcal E}k'{}^n +{\mathcal P}k'{}^m]\hat \delta u^\nu (k-k')\nonumber\\
&+&
u_0^\nu \int \frac{dk'}{(2\pi)^4}[{\mathcal E}k'{}^n +{\mathcal P}k'{}^m]\hat \delta u^\mu (k-k') \nonumber\\
&+&H_+^{\mu\nu}\hat \delta \rho(k)+
H_-^{\mu\nu}\hat\delta p(k)\,,
\eea
Again, to proceed further analytically we need (physically reasonable) assumptions 
on $\hat \delta u^\mu(k)$, $\hat \delta \rho(k)$
and $\hat \delta p(k)$. 
For instance, the simplest case is:
\beq
\label{simple_delta u}
\hat \delta u^\nu (k)=(2\pi)^4C_{\delta u}^\nu \delta (k)\,,
\eeq
with $C_{\delta u}^\nu $ a constant vector.

This implies
\bea
\hat S_1^{\mu\nu}(k)&=& (u_0^\mu C_{\delta u}^\nu+u_0^\nu C_{\delta u}^\mu) [{\mathcal E}k^n +{\mathcal P}k^m] \nonumber\\
&+&H_+^{\mu\nu}\hat \delta \rho(k)+
H_-^{\mu\nu}\hat\delta p(k)\,.
\eea
Assuming a power-law for $\hat \delta \rho(k)$  and $\hat\delta p(k)$ 
(e.g., $\hat \delta \rho(k)\sim C_{\delta \rho} k^p$), we are redirected exactly to 
the same  mathematical treatment as for the previous case of $S_0^{\mu\nu}$.

Let us now examine the case of $S_2^{\mu\nu}$ and limit our considerations to the first term,
\beq
S_{2a}^{\mu\nu}(x)=(\rho_0+p_0)R_{u, u}^{\mu\nu},
\eeq
since all the others can be treated similarly.
Passing to the Fourier space we find
\bea
\hat S_{2a}^{\mu\nu}(k)&=& \int \frac{d^4k_1}{(2\pi)^4}\frac{d^4k_2}{(2\pi)^4}(\hat \rho_0(k_1)+\hat p_0(k_1))\hat \delta u^\mu (k_2)\times\nonumber\\
&& \hat \delta u^\nu (k-k_1-k_2)\,.
\eea
Using  Eq. \eqref{hat_rho_p} the above expression becomes
\bea
\hat S_{2a}^{\mu\nu}(k)&=& \int \frac{d^4k_1}{(2\pi)^4}\frac{d^4k_2}{(2\pi)^4}({\mathcal E}k_1^n +{\mathcal P}k_1^m)\hat \delta u^\mu (k_2)\times\nonumber\\
&& \hat \delta u^\nu (k-k_1-k_2)\,,
\eea
and again to proceed further we need a physically reasonable expression for $\hat \delta u^\mu (k)$.
In the simple case \eqref{simple_delta u} we obtain
\bea
\hat S_{2a}^{\mu\nu}(k)&=& C_{\delta u}^\mu C_{\delta u}^\nu  \int  d^4k_1  d^4k_2 ({\mathcal E}k_1^n +{\mathcal P}k_1^m)\times \nonumber\\
&& \delta (k_2)\delta (k-k_1-k_2)\nonumber\\
&=&C_{\delta u}^\mu C_{\delta u}^\nu  ({\mathcal E}k^n +{\mathcal P}k^m)\,.
\eea
Correspondingly
\bea
\frac{h_{2a}{}_{\mu\nu}(x)}{16 \pi G}&=& -  C_{\delta u}^\mu C_{\delta u}^\nu \int \frac{d^4k}{(2\pi)^4} ({\mathcal E}k^{n-2} +{\mathcal P}k^{m-2}) e^{ik\cdot x}\nonumber\\
&=& -   C_{\delta u}^\mu C_{\delta u}^\nu [{\mathcal E}J_{n-2}(x)+{\mathcal P}J_{m-2}(x)]\nonumber\\
&=& -   C_{\delta u}^\mu C_{\delta u}^\nu \left[{\mathcal E}\frac{C_{n-2}}{x^{n+2}} +{\mathcal P}\frac{C_{m-2}}{x^{m+2}}\right]
\,.
\eea
 
Extending these considerations to the other terms entering $S_2^{\mu\nu}$ or to the remaining component  of the fluid source 
$S_3^{\mu\nu}$ is performed along the same lines as above. Namely, it is conceptually straightforward, albeit
a bit more involved from a mathematical standpoint.

\subsection{Timelike geodesics and particle scattering in fluctuating spacetime}

\begin{figure}[hb]
\centering
\includegraphics[scale=0.40]{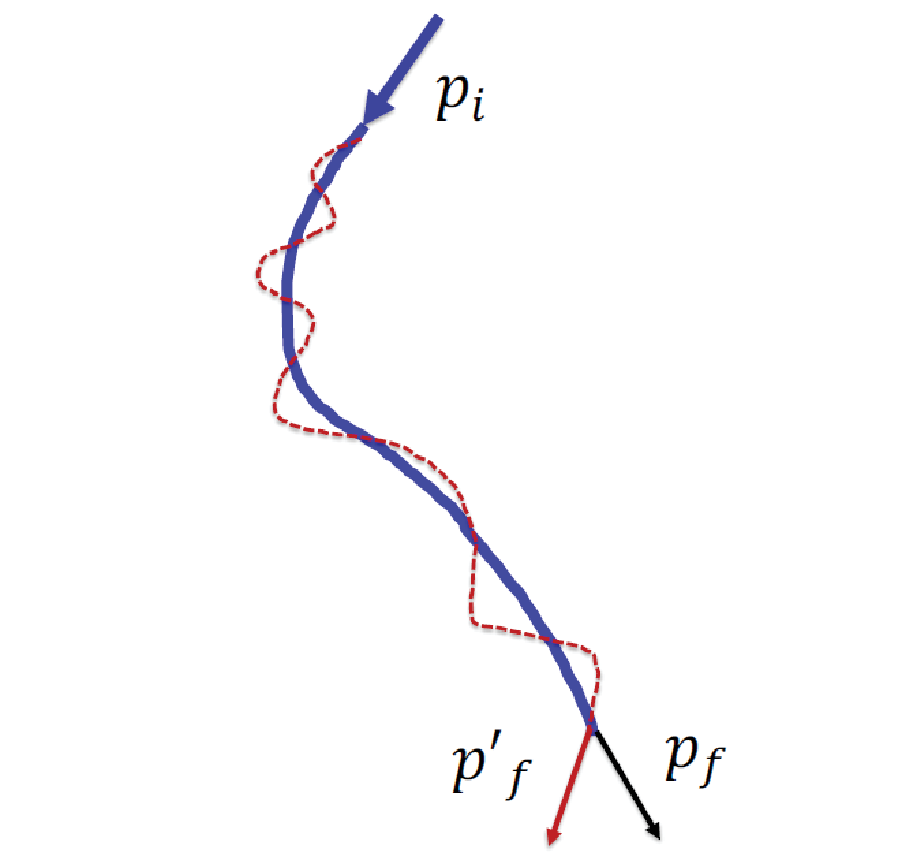}   
\caption{\label{fig:3}Geodesic motion of a free particle on a smooth (thick line) and
fluctuating (dashed)  metric manifold.  Subscripts ``i\rq\rq and ``f\rq\rq denote the
initial and final momenta. The effects of metric fluctuations can be paralleled to
a scattering process (similar to temperature fluctuations in classical fluids)
represented here as the departure of $p'_f$ from $p_f$.
}
\end{figure}

Armed with the above formalism,  we next  proceed to study the 
timelike geodesics (the orbits of massive particles with mass $m$) of the 
turbulence- perturbed metric 
$g^{\mu\nu}(x)=\eta^{\mu\nu}-h^{\mu\nu}(x)$, with unit tangent vector
\beq
p_\alpha=m \frac{dx^\alpha}{d\tau}=\frac{dx^\alpha}{d\sigma}\,,\qquad 
\eeq
such that
\beq
\frac{dp_\alpha}{d\sigma}=\frac12 \partial_\alpha h_{ {\mu\nu}}(x) p^\mu p^\nu\,,
\eeq
where $\sigma=\frac{\tau}{m}$ and $\tau$ the proper time parameter. The situation is schematically depicted in Fig. \ref{fig:3}.

It is also straightforward to derive the variation of the particle's 4-momentum 
when scattered by the (spacetime metric generated) fluid, namely  
\beq
\label{variation}
\Delta p_\alpha=\frac12 \int_{-\infty}^\infty d\sigma \partial_\alpha h_{\mu\nu}(x)\bigg|_{x=x(\sigma)} p^\mu p^\nu\,.
\eeq
Working at the first order in $G$, since $h_{\mu\nu}$ is already $O(G)$, all the other ingredients entering
the right-hand-side of Eq. \eqref{variation} can be replaced by their zeroth-order (free
motion in a flat spacetime) approximations, i.e., constant momenta 
($p^\alpha=p_-^\alpha$=constant)  and straight (incoming) world lines:
\beq
\label{geo_before}
x^\alpha(\sigma)=x^\alpha(0)+p_-^\alpha \sigma\equiv b^\alpha +p_-^\alpha \sigma\,,
\eeq
where we have denoted $x(0)=b$.
As a result:
\bea
\label{variation2}
\Delta p_\alpha&=&\frac12 p^\mu_- p^\nu_- \int_{-\infty}^\infty d\sigma \partial_\alpha h_{{\mu\nu}}(x) \bigg|_{x^\alpha=b^\alpha+p_-^\alpha \sigma}\nonumber\\
&=& -4 G (2\pi)^2  p^\mu_- p^\nu_- i 
\int \frac{d^4k}{(2\pi)^4}  \frac{\hat S_{\mu\nu}(k)}{k^2}k_\alpha \delta (k\cdot p_-) e^{ik\cdot b}  \,,\nonumber\\
\eea
with $\delta (k\cdot p_-)$ arising from the integration over $\sigma$. 

Here, we can take $\hat S_{\mu\nu}(k)=\hat S_0^{\mu\nu}(k)$ as given, for 
example, by Eq. \eqref{hat_S0_munu}, 
\beq
\hat S_{\mu\nu}(k)=H_{+\,\mu\nu}{\mathcal E}k^n +H_{-\,\mu\nu}{\mathcal P}k^m\,.
\eeq
Let us introduce the notation
\beq
H_\pm {}_{\mu\nu} p^\mu_- p^\nu_-=L_\pm={\rm const}\,.
\eeq
To perform this integral we choose a coordinate system such that 
the unperturbed particle moves along a straight line parallel to the $y$ axis, namely:
\beq
p_-=m(\gamma \partial_t -\sqrt{\gamma^2-1}\partial_y)\,,
\eeq
The parametric equations of the associated orbit then become
\bea
t(\sigma)&=& m\gamma \sigma\,,\qquad  x(\sigma)= b\,, \nonumber\\ 
y(\sigma)&=& -m \sqrt{\gamma^2-1} \sigma \,,\qquad
z(\sigma)= 0\,.
\eea
As a result, we obtain:
\begin{widetext}
\bea
\label{variation3}
\Delta p_\alpha&=&  -4 G  \frac{(2\pi)^2}{m\gamma} i \int \frac{dk_0 d^3k}{(2\pi)^4}[L_+ {\mathcal E}k^{n-2}+L_- {\mathcal P}k^{m-2}] k_\alpha e^{ik_x b}  \,
\delta(k_0-\frac{\sqrt{\gamma^2-1}}{\gamma}k_y)\nonumber\\
&=&  -4 G  \frac{(2\pi)^2}{m\gamma} i \int \frac{d^3k}{(2\pi)^4}[L_+ {\mathcal E}k^{n-2}+L_- {\mathcal P}k^{m-2}] k_\alpha e^{ik_x b}|_{k_0=\frac{\sqrt{\gamma^2-1}}{\gamma}k_y}\,,
\eea
with $k_0=\frac{\sqrt{\gamma^2-1}}{\gamma}k_y$ the on-shell condition and $d^3k=dk_x dk_y dk_z$. 

Therefore $k^2=k_x^2+\left(\frac{k_y}{\gamma}\right)^2+k_z^2\equiv k_\perp^2+k_z^2$ on-shell.
Finally
\bea
\Delta p_\alpha
&=&  -4 G  \frac{i}{m\gamma(2\pi)^2}  \left[L_+ {\mathcal E}\int  d^3k   k^{n-2}k_\alpha e^{ik_x b}+L_- {\mathcal P}\int  d^3k  k^{m-2}k_\alpha e^{ik_x b}\right]  \,,
\eea
\end{widetext}
On symmetry grounds, we have:
\beq
\Delta p_z=0\,,
\eeq
and,  because of the on-shell condition,  we obtain
\beq
\Delta p_0=\frac{\sqrt{\gamma^2-1}}{\gamma}\Delta p_y\,.
\eeq

Again for symmetry reasons (the integrand is an odd function of $k_y$), $\Delta p_y=0$ implying 
$\Delta p_0=0$,
\bea
\Delta p_y
&=&  - \frac{4iG}{m\gamma(2\pi)^2}  \left[L_+ {\mathcal E}\int  d^3k   k^{n-2}k_y e^{ik_x b}\right. \nonumber\\
&+&\left.L_- {\mathcal P}\int  d^3k   k^{m-2}k_y e^{ik_x b}\right]=0  \,.
\eea
Finally, $\Delta p_x$   can be expressed as a derivative with respect to $b$, 
\bea
\Delta p_x
&=& -  \frac{4  G}{m\gamma(2\pi)^2}\frac{\partial}{\partial b} \Psi \,,  
\eea
where
\bea
\Psi&=& L_+ {\mathcal E}\int  d^3k   k^{n-2} e^{ik_x b}
\nonumber\\&+& 
L_- {\mathcal P}\int  d^3k   k^{m-2} e^{ik_x b}\,.\qquad
\eea
In this case, the integration over $k_z$ becomes trivial by using the relation
\bea
\int dk_z k^n&=&\int dk_z (k_\perp^2+k_z^2)^{n/2}\nonumber\\
&=&k_\perp^{n+1}\sqrt{\pi}\frac{\Gamma(-\frac12 -\frac{n}{2})}{\Gamma(-\frac{n}{2})}\nonumber\\
&=&k_\perp^{n+1}B_n
\,.
\eea
Consequently
\bea
\Psi&=&
L_+ {\mathcal E}B_{n-2}  \int  dk_xdk_y  k_\perp^{n-1} e^{ik_x b}\nonumber\\
&+& L_- {\mathcal P}B_{m-2} \int  dk_xdk_y   k_\perp^{m-1}e^{ik_x b}\,,  
\eea
The basic integral to be computed  is then:
\bea
Y_{n}(b)&=&\int  dk_xdk_y  k_\perp^{n-1} e^{ik_x b}\nonumber\\
&=&\int  dk_xdk_y  \left(k_x^2+\left(\frac{k_y}{\gamma}\right)^2\right)^{(n-1)/2}  e^{ik_x b}\nonumber\\
&=&\gamma \int  dk_xdk_y  \left(k_x^2+  k_y^2\right)^{(n-1)/2}  e^{ik_x b}\,,
\eea
This yields:
\beq
\Psi=\left[L_+ {\mathcal E}B_{n-2}Y_n(b)   +L_- {\mathcal P}B_{m-2}Y_m(b)\right] \,.
\eeq
For $n<0$ ($n$ is assumed to be real)
\bea
Y_{n}(b)&=& \sqrt{\pi}\gamma \frac{\Gamma(-\frac{n}2)}{\Gamma( \frac{1-n}2)} \int_{-\infty}^\infty  dk_x    k_x^n  e^{ik_x b}\nonumber\\
&=&
 \sqrt{\pi}\gamma \frac{\Gamma(-\frac{n}2)}{\Gamma( \frac{1-n}2)}\frac{1}{b^{n+1}}\int_{-\infty}^\infty  du    u^n  e^{iu}
\,,
\eea
For $n=-1$ the above rescaled expression cannot be used. 
Working directly with the un-rescaled expression, we have
\beq
Y_{-1}(b)=  \pi \gamma   \int  dk_x    \frac{e^{ik_x b}}{k_x}   \,,
\eeq
implying the following Dirac-delta result
\beq
\frac{d}{db}Y_{-1}(b)=2i\pi^2\gamma \delta(b) \,.
\eeq
Other values of $n$ are also of interest; for $n=-3$, we compute:
\bea
Y_{-3}(b)&=& \sqrt{\pi}\gamma \frac{\Gamma(\frac{3}2)}{\Gamma(2)}b^2\int_{-\infty}^\infty  du    u^{-3}  e^{iu}   \nonumber\\
&=&  \frac12 \pi \gamma  b^2 i\int_{-\infty}^\infty  du     \frac{\sin u}{u^3}\nonumber\\
&=& -\frac{\pi^2}{4} \gamma  b^2 i\,,
\eea
where the last integral diverges at $u=0$ and it has been evaluated by using the Partie Finie.
For $n=-5/3$ we find instead:
\bea
Y_{-5/3}(b)&=& \sqrt{\pi}\gamma \frac{\Gamma( \frac{5}6)}{\Gamma(\frac43)}b^{2/3} \int_{-\infty}^\infty  du    u^{-5/3}  e^{iu}\nonumber\\
&=& \sqrt{\pi}\gamma \frac{\Gamma( \frac{5}6)}{\Gamma(\frac43)}b^{2/3} i 2\int_{0}^\infty  du  \frac{\sin u}{u^{5/3}} \nonumber\\
&=&  3 \pi^{3/2}\gamma \frac{\Gamma( \frac{5}6)}{\Gamma(\frac43)\Gamma(\frac23)}b^{2/3} i\,,  
\eea
where
\beq
\Gamma\left(\frac43\right)\Gamma\left(\frac23\right)=\frac29 \pi \sqrt{3}\,,\qquad \Gamma\left(\frac56\right)\approx 1.1288\,.
\eeq
Therefore
\bea
Y_{-5/3}(b)
&=&  27 \pi^{1/2}\gamma \frac{\Gamma\left( \frac{5}6\right)}{2 \sqrt{3}}b^{2/3} i\nonumber\\
 &\approx&   15.5941 \gamma b^{2/3} i\,.
\eea
Summarizing, in the case $n=-1$ we end up with a $\Delta p_x \propto \delta (b)$, namely a 
particle scattered by a fluid with an energy spectrum of the type $\sim k^{-1}$  
experiences an instantaneous variation of its (initially constant) linear momentum. 
In the case of a energy spectrum of the type $\sim k^{n}$, we find a 
corresponding power law for the variation of the linear momentum,
$\Delta p_x \propto b^{-n-1}$. This may permit to discriminate between various 
equations of state of the fluid (including a fluid undergoing a turbulent behavior) 
by examining the variation of linear momentum for a particle  scattered by the fluid itself.    

 \subsection{Dissipative effects: viscosity and heat conduction}

As discussed in Sec. II, the Kolmogorov cascade is terminated by dissipative
effects; as a result it is of interest to extend our analysis to the case of a viscous fluid with 
heat conduction, as discussed in Ex. 22.7 of Ref. \cite{Misner:1973prb}.

To this purpose,  we increment Eq. \eqref{en_mom} with additional dissipative components: 
\bea
\label{en_mom}
T_{\rm visc}^{\mu\nu}&=& -\zeta \Theta(u) P(u)^{\mu\nu}-2\eta \sigma(u)^{\mu\nu}\,,\nonumber\\
T_{\rm visc}&=& -3\zeta \Theta(u)\,,\qquad\nonumber\\
T_{\rm heat}^{\mu\nu}&=& 2 u^{(\mu} q^{\nu)}\,,\qquad q\cdot u=0\,,\nonumber\\ 
T_{\rm heat}&=&0\,,
\eea
where $P(u)_{\alpha\beta}=g_{\alpha\beta}+u_\alpha u_\beta$ projects orthogonally to $u$, $\sigma(u)^{\alpha \beta}={\rm STF}_{\alpha\beta}[P(u)_{\alpha}^\mu P(u)_{\beta}^\nu \nabla_{\mu}u_\nu]$ is the symmetric and tracefree (STF) shear tensor of the fluid and $\Theta(u)=\nabla_\alpha u^\alpha$ is the expansion scalar. Moreover, $\zeta\ge 0$ denotes the coefficient of bulk 
viscosity, $\eta\ge 0$ the coefficient of dynamical viscosity and
\beq
q^\alpha=-\kappa P(u)^{\alpha\beta}[T_\beta+Ta(u)_\beta]\,,
\eeq
is the Eckart\cite{Eckart:1940zz}  law of conduction heat (later developed by C. Cattaneo \cite{Cattaneo}) 
with $\kappa$ the (constant) coefficient of thermal conductivity and 
$a(u)_\alpha=\nabla_u u_\alpha$ the acceleration of the fluid world lines.

It proves expedient to introduce the dissipative tensors
$S_{\rm visc}^{\mu\nu}=T_{\rm visc}^{\mu\nu}-\frac12 T_{\rm visc}g^{\mu\nu}$ and $S_{\rm heat}^{\mu\nu}=T_{\rm heat}^{\mu\nu}$ and consider their representation in the Fourier space. 
Assuming again $u^\mu=u_0^\mu=$constant in a Minkowski (flat) spacetime 
referred to Cartesian coordinates $\nabla_\alpha u_\beta=0$, 
hence $T_{\rm visc}^{\mu\nu}=0$ (i.e., a viscosity contribution appears 
at higher orders of the PM procedure), we obtain:
\bea
S_{\rm heat}^{\mu\nu}&=& u_0^{\mu} q^{\nu}+u_0^{\nu} q^{\mu}\nonumber\\
&=&-\kappa (u_0^\mu P(u_0)^{\nu\beta}+u_0^\nu P(u_0)^{\mu\beta})T_\beta\nonumber\\
&=&-\kappa (u_0^{ \mu} T^{\nu}+u_0^{ \nu} T^{\mu}+2u_0^{\mu} u_0^\nu u_0^\sigma T_\sigma)\,,
\eea
where $T^\mu=T^{,\mu}$ denotes the temperature gradient.

Moving to Fourier space:
\bea
\hat S_{\rm heat}^{\mu\nu}(k)&=&-\kappa[ 2 u_0^{( \mu} \tau^{\nu)} (k)
+2 u_0^{\mu} u_0^\nu u_0^\sigma \tau_\sigma (k)]  \,,\qquad
\eea
where
\beq
\tau^\nu (k)=\int d^4x e^{-ik\cdot x} T^{\nu}=ik^\nu \hat T(k)\,,
\eeq
and
\beq
\hat T(k)=\int  d^4x e^{-ik\cdot x} T(x)
\eeq
is the Fourier transform of the temperature.

Finally, we find
\bea
\hat S_{\rm heat}^{\mu\nu}(k)&=&-i\kappa \hat T(k)[ 2 u_0^{( \mu} k^{\nu)} 
+2 u_0^{\mu} u_0^\nu u_0^\sigma k_\sigma ]  \,.\qquad
\eea
To proceed further,  consistently with the present analysis,  
we consider a power-law spectrum 
\beq
\hat T(k)=T_0 k^\ell\,,
\eeq
for some power exponent $\ell$. 

This takes us back to the same basic integrals encountered in the non-dissipative treatment, 
without adding any further layer of mathematical complexity,  at least at 
the 1PM level of approximation. 

Future work is however needed to investigate the consequences of releasing
the main simplifying  assumptions,  such as  $u^\mu=u_0^\mu=$ constant.
Work along these lines is currently in progress.

\subsection{Nonlocal effects and link to fractional calculus}

Finally, we observe that the integral \eqref{J_q_int} is conducive to fractional 
calculus,   and particularly to a fractional version of the D'Alembert 
operator.  As is well known,  fractional derivatives 
describe nonlocal effects in space an time,  which is consistent
with the nature of turbulence \cite{FRACT,Nottale:2009azf,FRAC2}.  
Indeed, coherent structures display a finite lifetime,  meaning
that by the time their effects are felt in a given spacetime location, 
they have already or moved elsewhere in the fluid,  or possibly dissipated away,
whence the memory effect in both space and time.

It is very plausible to speculate that such form of dynamic 
memory should be enhanced by coupling to the gravitational 
degrees of freedom,  since the 
latter are driven by curvature,  itself a source
of nonlocality even in non-relativistic physics 
(think of the Poisson equation in electrostatics).

Clearly,  the 1PM framework falls short of capturing the interaction of turbulence with 
black-holes,  but, as mentioned above,  it is plausible to speculate that the 
emergence of nonlocal effects would only be strengthened in 
the presence of strongly curved spacetimes. 

\section{Concluding remarks}

We have inspected the perturbative effects of fluid turbulence on 
the gravitational metric and viceversa. 
Based on purely statistical steady-state scaling arguments, we have first studied the
qualitative viability of gravitational interference on the turbulent energy cascade.
Next, we have performed a detailed dynamic analysis within the simplified
framework of first-order Post-Minkowskian (1PM) gravity.
Despite not being analytically solvable and far from the strong-curvature 
conditions characterizing black-hole physics,  the 1PM analysis strongly hints
at the onset of turbulence-driven non-local effects on spacetime evolution.
In fact, it permits to pin down the real-spacetime scaling exponents of the 
perturbed metric as a function of the spectral exponents of turbulence.  

Although firm conclusions are necessarily hinging on a more quantitative
non-perturbative analysis, most likely by numerical means,  it is plausible to speculate that such
turbulence-driven nonlocal effects would only be accrued in the presence
of strongly curved spacetimes.

\appendix

\section{Evaluating the integral $J_q(x)$}
\label{app:Jq}

We start from the integral
\beq
J_q(x)=\int \frac{d^4 k}{(2\pi)^4}k^q e^{ik \cdot x}\,,
\eeq
where $x=(t,x^i)$ is a spacetime vector, $x^2=-t^2+{\mathbf x}^2$,
\beq
k=-\omega dt +K_i dx^i\,,\qquad K=\sqrt{\delta_{ij}K^i K^j}\,,
\eeq
(having denoted $k_0=-\omega$, $k_i=K_i$ for convenience) and
\beq
k^q =(-\omega^2+K^2)^{q/2}\,.
\eeq
For negative values of $q$ this integral is divergent along the lines $\omega=\pm K$ and should be regularized, see below.
Let us first reduce this integral as follows.
Using spherical coordinates in the 3-space of ${\mathbf K}$, we find
\beq
d^4 k =d\omega d^3K=d\omega \, K^2dK \, \sin \theta d\theta d\phi
\eeq
and
\beq
\label{Jq_def_integr}
J_q(x)=\int \frac{d^3 K}{(2\pi)^3} e^{iK_ix^i}A(t; K,q)\,.
\eeq
having defined
\beq
\label{eq:A_t}
A(t; K,q)=\int \frac{d\omega}{2\pi}e^{-i\omega t}(-\omega^2+K^2)^{q/2}\,.
\eeq

To compute the integral  \eqref{Jq_def_integr} let us
 assume that ${\mathbf K}$ is aligned with the $z$ axis of the spherical coordinates so that
\beq
K_ix^i=K |{\mathbf x}| \cos \theta\,,
\eeq
and (after trivial integration over $\phi$)
\beq
J_q(x)=\int_0^\infty \frac{K^2dK \, }{(2\pi)^2}A(t; K,q) \int_0^\pi \sin \theta d\theta e^{iK |{\mathbf x}| \cos \theta}\,.
\eeq
The integral in $\theta$ can be easily performed
\beq
\int_0^\pi \sin \theta d\theta e^{iK |{\mathbf x}| \cos \theta}=2\frac{\sin (K |{\mathbf x}|) }{K |{\mathbf x}|}\,,
\eeq
and therefore
\beq
J_q(x)=\frac{1}{2\pi^2 |{\mathbf x}|}\int_0^\infty  K dK  A(t; K,q)  \sin (K |{\mathbf x}|)  \,.
\eeq
Under assumptions $-2<q<0$ ($q$ real), $t\not =0$, $|{\mathbf x}|\to |{\mathbf x}|-i\epsilon$ (with $\epsilon>0$, later sent to 0) we find
\beq
\label{J_q_Mink}
J_q(x)= i \frac{2^q}{\pi^2 x^{4+q}}\frac{\Gamma(1+\frac{q}{2})}{\Gamma(-(1+\frac{q}{2}))}\equiv \frac{i C_q}{x^{4+q}}\,,
\eeq
with $C_q$ given in Eq. \eqref{J_q_and_C_q} above and repeated below for convenience,
\beq
\label{C_q_def_2}
C_q=\frac{2^q}{\pi^2}\frac{\Gamma(1+\frac{q}{2})}{\Gamma(-(1+\frac{q}{2}))}\,.
\eeq
The Minkowskian (singular) integral leads to a complex quantity, meaning  that
a Wick-rotation is required to turn it into a real,  hence to a physically acceptable metric.

To show this in detail, we repeat  the previous computation in the 
Euclidean case, by using   Eqs. 3 and 5 of Ref.  \cite{Halliday:1987an}, conveniently 
rewritten here as    
\bea
I_{m,0,d}&=&\int d^dk (k^2)^m =(-1)^m \pi^{d/2}\Gamma(m+1)\delta_{m+d/2, 0}\,,\nonumber\\
I_{m,n,d}&=&\int d^dk (k^2)^m (2k\cdot x)^n \nonumber\\
&=&\frac{m!n!}{(n/2)!}\pi^{d/2}(-1)^m (x^2)^{n/2}  \delta_{m+(d+n)/2, 0}\,,
\eea
when $d=4$.
The result is:
\bea
\label{J_q_Euclid}
J_q^{\rm Wick}(x)&=&\int \frac{d^4 k}{(2\pi)^4}k^q e^{ik \cdot x}\nonumber\\
&=&\sum_{n=0}^\infty\frac{i^n}{n!(2\pi)^4 2^n}\int  d^4 k (\kappa^2)^{q/2} (2k \cdot x)^n\nonumber\\
&=& \sum_{n=0}^\infty\frac{i^{n+q}}{n!(2\pi)^4 2^n}\frac{(q/2)!n!}{(n/2)!}\pi^2  (x^2)^{n/2}  \delta_{(q+n)/2+2, 0}\nonumber\\
&=& \frac{C_q}{x^{q+4}} \,,
\eea
where $C_q$ is given in Eq. \eqref{C_q_def_2} and $x=x_{\rm Euclidean}$.

We see indeed that $J_q^{\rm Wick}(x)$,  Eq. \eqref{J_q_Euclid}, differs from its 
Minkowskian analogue, Eq. \eqref{J_q_Mink}, by the imaginary unit, consistently with
 the Wick rotation performed to regularize it 
(see also Ref.  \cite{Halliday:1987an} for the convergence conditions controlling
 the validity of this result). 
 
 However, the spacetime distance in Eq. \eqref{J_q_Euclid} is  Euclidean, and 
 extending the result to the Minkowskian case raises a question 
 whenever $x^2<0$ (timelike distance). 
 
A possible solution is to reinstate positive-definetess is to 
introduce  the absolute value,
\beq
x^2_{\rm Euclidean}=t^2+x^2=|t^2+{\mathbf x}^2|\to |(it)^2+{\mathbf x}^2|=|-t^2+{\mathbf x}^2|\,;
\eeq
the latter leaves the Euclidean distance unaffected but plays the desired goal 
of removing ambiguities in the Minkowskian case.

However,  at this stage the above fix and the subsequent discussion is purely empirical,  indicating  that 
further detailed investigation is required to fully unveil the
nature of the turbulence-driven metric singularities in the Minkowskian case.

\section*{Acknowledgments}
DB thanks P. Mastrolia for useful discussions and ICRANet for partial support. 
DB acknowledges sponsorship of the Italian Gruppo Nazionale 
per la Fisica Matematica (GNFM) of the Istituto Nazionale di Alta Matematica (INDAM).
SS kindly acknowledges funding from the European Research
Council under the Horizon 2020 Programme Grant Agreement n. 739964 (\lq\lq COPMAT").


\end{document}